\documentclass[amsmath,reprint,amssymb,aps, longbibliography]{revtex4-1} 
\usepackage{graphicx}
\usepackage{dcolumn}
\usepackage{bm,times}
\usepackage[colorlinks,citecolor=blue,linkcolor=red]{hyperref} 
\usepackage{color}  

\begin{document}

\preprint{APS/123-QED}

\title{Universal Relations of Energy Flow, Acoustic Spin and Torque for Near-Field Acoustic Tweezers }

\author{Yang Long}
\author{Chenwen Yang}
\author{Hong Chen}
\author{Jie Ren}
\email{Corresponding author: xonics@tongji.edu.cn}
\affiliation{%
Center for Phononics and Thermal Energy Science, China-EU Joint Lab On Nanophononics, Shanghai Key Laboratory of Special Artificial Microstructure Materials and Technology, School of Physics Science and Engineering, Tongji University, Shanghai 200092, China
}%

\date{\today}

\begin{abstract}
Acoustic spin, radiation torque, energy flow, and reactive power are of significant importance from both fundamental and practical aspects, responsible for flexible tweezer manipulations and near-field sound directionality.
Nevertheless, the intrinsic relations among these physical quantities are far from clear.
Here, we prove the universal geometric relations among them in acoustics, independent on wave structure details. Particularly, we connect acoustic spin and torque to the cross product of  time-averaged energy flow and reactive power, as well as to the local vorticity of energy flow. These relations are universally valid, verified in a variety of different acoustic systems.
We also demonstrate the multipole mechanical torques and forces generated in three acoustic near-field sources: Janus, Huygens and Spin sources, applying on small lossy particles.
These universal geometric relations uncover hidden locking relations beyond simple spin-momentum locking of near-field waves, and show the basic principles between the acoustic spin, radiation torque, and energy flow, reactive power. 
\end{abstract}
\maketitle

{\bf Introduction} 

Since the pioneering works by J. H. Poynting, radiation force and energy flow are studied for many decades, which are playing fundamental roles in optical and acoustic tweezers. Meanwhile, the spin angular momentum (SAM) in the near-field wave physics has attracted much attention recently: non-trivial topological properties for surface optical waves~\cite{bliokh2015quantum, bliokh2019topological, bliokh2014extraordinary}, selective optical excitations~\cite{bliokh2015spin, aiello2015transverse}, optical torque~\cite{ bliokh2015transverse, bekshaev2015transverse}, and chiral quantum interaction enhancement~\cite{lodahl2017chiral}.
The SAMs in acoustic and elastic waves have also been theoretically proposed~\cite{long2018intrinsic} and experimentally verified~\cite{shi2019observation} in parallel, following with applications, $i.e.$ momentum-locked transverse SAM in surface evanescent waves~\cite{shi2019observation, long2018intrinsic, bliokh2019transverse} or mirror-symmetry breaking waveguide~\cite{Long2020}, multiphysics interactions between surface phononic waves and magnons~\cite{Zhang2020} and self-rotation torques on lossy particles~\cite{shi2019observation, toftul2019acoustic}. 
As unveiled in Refs.~\cite{picardi2018janus, long2020symmetry, Wei2020, Long2020a}, there are another two important physical quantities in near-field waves, time-averaged energy flow $\bm{\mathcal{J}}$ and reactive power $\bm{\mathcal{R}}$, which exhibit inner symmetric relations in evanescent wave modes, not only for acoustics~\cite{long2020symmetry, Wei2020} but also for optics~\cite{picardi2018janus,  Long2020a}, resulting in fertile selective near-field excitations beyond spin-momentum locking~\cite{long2020symmetry, Long2020a}.
These research works have suggested 
possible connections among the SAM, $\bm{\mathcal{J}}$ and $\bm{\mathcal{R}}$ in near-field waves, which however are still not clear, and need to be uncovered. 

In this work, we unveil the universal geometric relations in near-field acoustic SAM and show the multipole mechanical torques fo near-field selective sources. We find that the acoustic SAM density $\bm{s}$ and torques can be directly related with the time-averaged energy flow $\bm{\mathcal{J}}$ and reactive power vector $\bm{\mathcal{R}}$ as: $\bm{s} \propto \bm{\mathcal{J}}\times\bm{\mathcal{R}}$ with right(left) handedness in right (left) handed medium.
The acoustic SAM and torque are also related to the local vorticity of energy flow $\nabla\times\bm{\mathcal{J}}$, resulting in universal existence conditions and geometric features for acoustic SAM. 
We verify the relations in diverse systems, including surface acoustic waves, surface Bessel waves, near-field acoustic sources, Bessel and Laguerre- Gaussian beams with non-zero topological charges and so on.
These relations show hidden locking relations beyond simple spin-momentum locking of near-field waves, and show the basic principles between the acoustic spin, radiation torque, and energy flow, reactive power. 
Furthermore, the multipole mechanical operations for various near-field acoustic sources have been demonstrated. 
Our work would pave the way about the understanding of geometric features of near-field wave physics and could inspire new insights for acoustic tweezer manipulations.

{\bf Results}

{\bf Universal relations between acoustic spin, energy flow, and reactive power.} 
The acoustic spin density is associated with the local acoustic velocity field as $\bm{s} = \frac{\rho}{2\omega} {\rm Im}[\bm{v}^*\times\bm{v}]$~\cite{shi2019observation, long2018intrinsic}, where $\rho$ is the mass density of medium, $\omega$ is the frequency, ${\rm Im}[\cdot]$ is the imaginary part and $\bm{v}$ is the acoustic velocity vector field. Obviously, the nonzero $\bm{s}$ will correspond to the circularly (or generally elliptical) polarized velocity field, which makes $\bm{s}$ a local property of the velocity field.  Besides the SAM, there are two other important physical quantities in near-field waves that can reflect 
flow directions and field confinements for acoustic waves~\cite{long2020symmetry}: the time-averaged energy flow $\bm{\mathcal{J}}=\frac{1}{2} {\rm Re}[p^*\bm{v}]$ and the reactive power vector $\bm{\mathcal{R}} = \frac{1}{2} {\rm Im}[p^*\bm{v}]$, where $p$ is the pressure field and ${\rm Re}[\cdot]$(${\rm Im}[\cdot]$) is the real(imaginary) part.  From acoustic equations, the reactive power vector can be represented as: $\bm{\mathcal{R}} = - \frac{1}{4\rho\omega} \nabla |p|^2 $. Clearly, the physical meaning of $\bm{\mathcal{R}}$ can be understood as the non-zero gradient of the potential energy from the pressure field, so that of curl-free nature  $\nabla\times\bm{\mathcal{R}}=0$.

According to the definitions of $\bm{\mathcal{J}}$ and $\bm{\mathcal{R}}$, ${\rm Re}[p^*\bm{v}] = \frac{1}{2}(p^*\bm{v} + p\bm{v}^*)$, and ${\rm Im}[p^*\bm{v}] = \frac{1}{2i} (p^*\bm{v} - p\bm{v}^*)$, we obtain that $\bm{\mathcal{J}}\times\bm{\mathcal{R}} = \frac{|p|^2}{8} \frac{\bm{v}^*\times\bm{v}-\bm{v}\times\bm{v}^*}{2i}$.
Considering further ${\rm Im}[\bm{v}^*\times\bm{v}] = \frac{1}{2i} (\bm{v}^*\times\bm{v} - \bm{v}\times\bm{v}^*)$ and the definition of acoustic SAM density, we finally find a universal geometric relation about the acoustic SAM as  (More details in Supplementary information):
\begin{equation}
\bm{s} = \frac{4 \rho}{\omega |p|^2} \bm{\mathcal{J}}\times\bm{\mathcal{R}}.
\label{eq:s}
\end{equation}
From this cross product relation, we can see that although the acoustic spin is physically defined by local velocities~\cite{shi2019observation, long2018intrinsic, bliokh2019transverse}, it can be determined simultaneously by time-averaged energy flows and spatially inhomogeneous field distributions. For given $\bm{\mathcal{J}}$ and $\bm{\mathcal{R}}$ in acoustic waves, $\bm{s}$ becomes determinate and does not have inner degree of freedom like Jones vector in transverse planes for polarized optical waves. The physical reason of the locking relation between $\bm{s}$ and $\bm{\mathcal{R}}$ can be understood as imperfect cancellations of neighbour spin momentum loops in spatially inhomogeneous wave fields~\cite{bliokh2014extraordinary, long2018intrinsic}. 

Further, considering $\nabla\times (p^*\bm{v}) = \nabla p^* \times \bm{v}$ since $\nabla\times\bm{v} =0$ for the longitudinal wave, we can also obtain another universal geometric relation as:
\begin{equation}
\bm{s} = \frac{1}{\omega^2} \nabla\times\bm{\mathcal{J}},
\label{eq:sJ}
\end{equation}
which reflects that the divergence-free $\bm{s}$ can be connected to the non-zero local vorticity of $\bm{\mathcal{J}}$, relating to the local circulation of energy flow in the zero loop limit. Combined with the universal geometric relations, Eq.~(\ref{eq:s}) and Eq.~(\ref{eq:sJ}), it is readily to see that the inhomogeneous potential energy gradient $\bm{\mathcal{R}}$ will lead to finite energy flow vorticity $\nabla\times\bm{\mathcal{J}}$ for acoustics.
In the following, we will demonstrate several examples to improve understandings and show the validity of the universal geometric relations.

\begin{figure}[tp!]
\centering
\includegraphics[width=\linewidth]{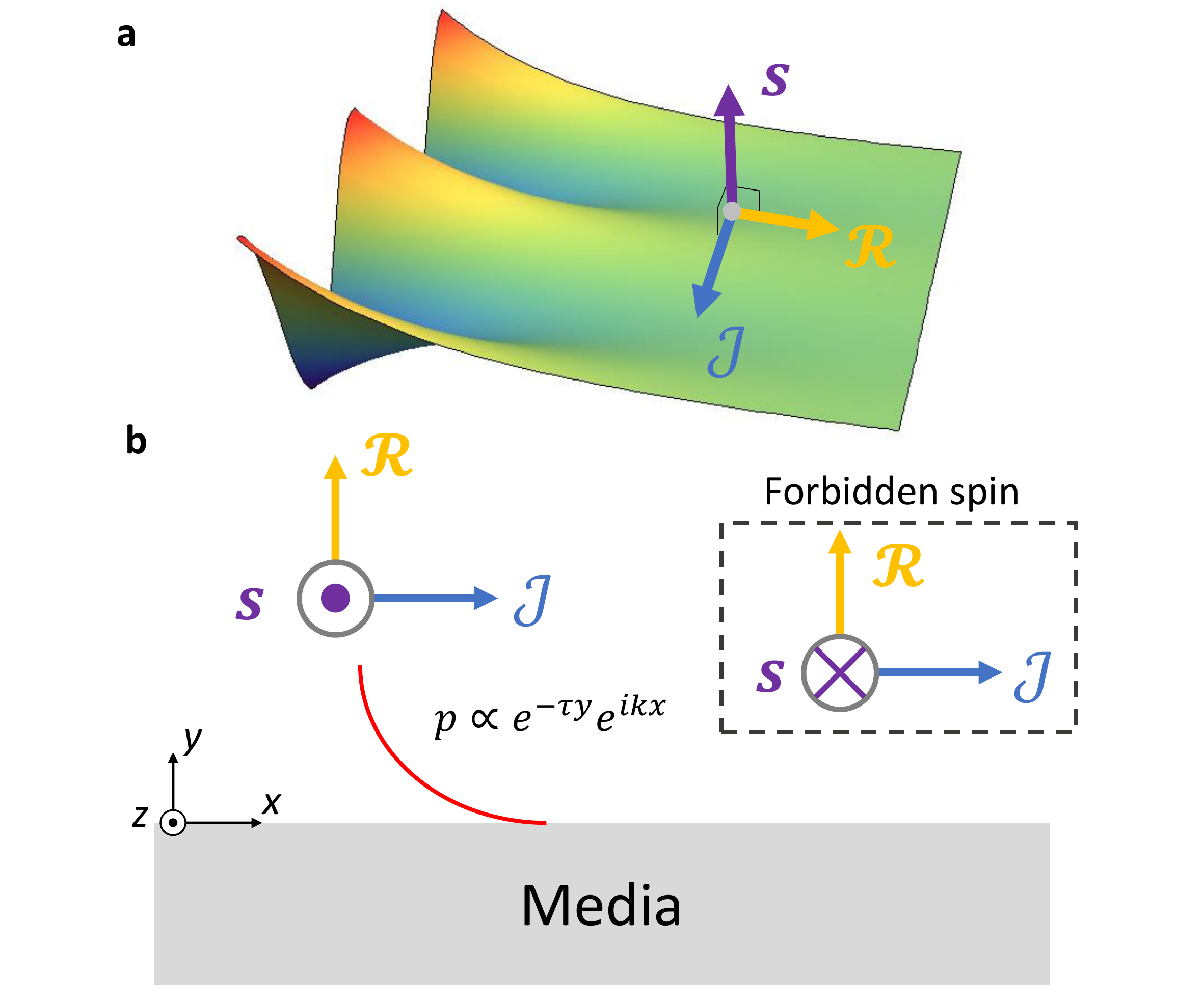}
\caption{{\bf Universal geometric features in acoustic surface evanescent waves.} \textbf{a} The directions of the time-averaged energy flow $\bm{\mathcal{J}}$ and reactive power $\bm{\mathcal{R}}$ as the spatially gradient of the acoustic field will determine the acoustic SAM density $\bm{s}$. \textbf{b} Although the surface modes have the spin-momentum ($\bm{s}-\bm{\mathcal{J}}$) locking relation, some spin for modes will be forbidden due to the extra locking of $\bm{s}-\bm{\mathcal{R}}$.}
\label{fig:evanescent}
\end{figure}

\begin{figure}
\centering
\includegraphics[width=\linewidth]{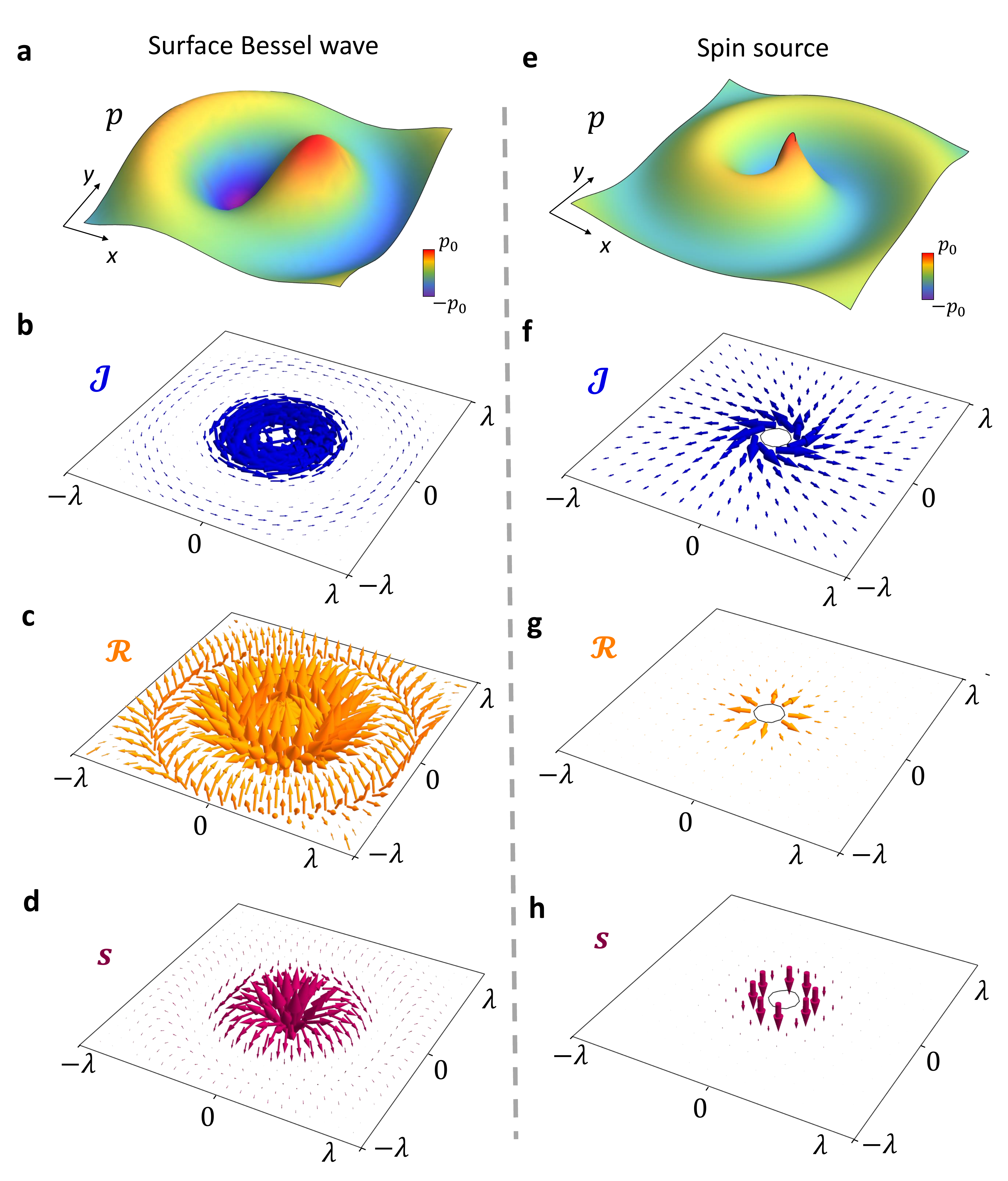}
\caption{{\bf Universal geometric features in acoustic surface Bessel waves and acoustic spin source  $D_x+i D_y$}.  Their pressure field $p$ (\textbf{a,e}) that is normalized by the maximum amplitude $p_0$, time-averaged energy flow $\bm{\mathcal{J}}$ (\textbf{b,f}), reactive power $\bm{\mathcal{R}}$ (\textbf{c,g}) and SAM density $\bm{s}$ (\textbf{d,h}) have been demonstrated. Here, for surface Bessel waves, $k = 1.2 k_0$, $\ell = 1$, $\lambda$ is the wavelength. For the spin source, the region around the spin source with the distance smaller than $\lambda/10$ (the center black circle) is not shown for avoiding the singularity of the field, $D_x$ and $D_y$ denote the acoustic dipoles along $x$ and $y$ direction, respectively.}
\label{fig:spinsource}
\end{figure}

{\bf Verification of universal relations in diverse acoustic wave structures.} 
The first famous spatially confined propagating near-field wave should be surface evanescent wave~\cite{shi2019observation, long2018intrinsic, bliokh2019transverse}. The SAM and spin-momentum locking of acoustic surface evanescent wave have been observed and verified recently~\cite{shi2019observation, long2020symmetry}. 
The physical reason behind spin-momentum locking in acoustic evanescent waves was considered as the analogy of the quantum spin Hall effects with non-zero spin Chern number~\cite{bliokh2015quantum, shi2019observation, long2018intrinsic} and described by topological properties of four-momentum operators for acoustic surface modes~\cite{bliokh2019klein}. 
Without the topological argument, the acoustic spin-momentum locking can be naturally understood with the help of the universal geometric relations. Considering the evanescent wave form $p = p_0 e^{- \tau y}e^{i (kx-\omega t)}$,  where $\tau=\sqrt{k^2 - k_0^2}$, $k_0 = \omega/c$, $c$ is the sound speed in the air, $y=0$ is the surface position, we can obtain that $\bm{\mathcal{J}} = \frac{|p_0|^2}{2\rho\omega} k e^{-2\tau y} \bm{e}_z$, $\bm{\mathcal{R}} = \frac{|p_0|^2}{2\rho\omega} \tau e^{-2\tau y} \bm{e}_y$ and $\bm{s} = \frac{|p_0|^2}{\rho\omega^3} k\tau e^{-2\tau y} \bm{e}_z$, as shown in Fig.~\ref{fig:evanescent}. It is readily to check that they satisfy the universal geometric relations of Eq.~(\ref{eq:s}) and Eq.~(\ref{eq:sJ}). Although spin-momentum locking reflects that the sign of SAM is strongly locked with $k$~\cite{bliokh2015quantum, van2016universal}, it can not tell us which sign of the SAM will be locked with the given $k$, but Eq.~(\ref{eq:s}) does. As unveiled in Fig.~\ref{fig:evanescent}, it is physically impossible for surface modes with $k>0$ to carry the SAM $\bm{s}<0$. The Eq.~(\ref{eq:s}) can be regarded as a hidden geometric property that restricts and forbids spin-momentum pairs in topological surface modes~\cite{shi2019observation, bliokh2019klein}. It is worth noting that although Eq.~(\ref{eq:s}) indicates the right-hand relation of $\bm{\mathcal{J}}$, $\bm{\mathcal{R}}$ and $\bm{s}$, the left-hand relation (so that the forbidden spin) would be possible in effective left-handed medium, $i.e.$, the double-negative acoustic metamaterial with negative mass density and negative compressibility simultaneously.  
The geometric relation of $\bm{\mathcal{J}}$, $\bm{\mathcal{R}}$ and $\bm{s}$ in near-field waves described by Fig.~\ref{fig:evanescent} is verified by the experimental data observed in Ref.~\cite{long2020symmetry}. 

The second example is the acoustic surface wave in the cylindrical coordination. The acoustic surface Bessel waves $p=p_0 J_{\ell}(k r)e^{i\ell \varphi} e^{-\tau z} e^{-i\omega t}$ in Fig.~\ref{fig:spinsource}(a) will carry non-zero SAM, where $J_{\ell}(k r)$ is the Bessel function of the first kind with order $\ell$, $\tau =\sqrt{k^2 - k_0^2}$ and $(r,\varphi,z)$ are the cylindrical coordinates in real space. One can obtain  $\bm{\mathcal{J}}=\frac{|p_0|^2}{2\rho \omega} \frac{\ell}{r} J_{\ell}^2(k r) e^{-2\tau z} \bm{e}_\varphi$ in Fig.~\ref{fig:spinsource}(b). Its Bessel-type inhomogeneous surface acoustic field will induce non-zero spatially gradient of acoustic field $\nabla |p|^2 = J_{\ell}^2(k r) e^{-2\tau z}$, leading to nonzero $\bm{\mathcal{R}}$, as shown in Fig.~\ref{fig:spinsource}(c), such that $\bm{\mathcal{R}} =\frac{|p_0|^2}{2\rho \omega} e^{-2\tau z} (-\frac{k}{2}J_{\ell}(k r) (J_{\ell-1}(k r) - J_{\ell+1}(k r))\bm{e}_r + \tau J^2_{\ell}(k r)\bm{e}_z) $. According to the velocity fields, we can get $\ell$-dependent SAM as: $\bm{s} = \frac{|p_0|^2}{\rho \omega^3} \frac{\ell}{r} e^{-2\tau z} (\tau J_{\ell}^2(k r)\bm{e}_r +\frac{k}{2} J_{\ell}(k r)(J_{\ell-1}(k r)-J_{\ell+1}(k r))\bm{e}_z)$ in Fig.~\ref{fig:spinsource}(d). Clearly, they satisfy the universal geometric relations of Eq.~(\ref{eq:s}) and Eq.~(\ref{eq:sJ}). 

The third example is the setup of near-field acoustic sources that have been recently proposed and verified experimentally~\cite{long2020symmetry, Wei2020}, which can excite the evanescent modes selectively according to the geometric and symmetric features of the source itself. As a demonstration illustrated in Fig.~\ref{fig:spinsource}(e-h), the acoustic spin source that can selectively excite surface waves due to spin-momentum locking~\cite{long2020symmetry,shi2019observation, bliokh2019transverse} has exhibited $\bm{\mathcal{J}}$ (f), $\bm{\mathcal{R}}$ (e), $\bm{s}$ (g) in Fig.~\ref{fig:spinsource}, around its near-field subwavelength regions (smaller than $\lambda$). We check that these quantities around the near-field directional source also satisfy the relations of Eqs.~(\ref{eq:s},~\ref{eq:sJ}).
Furthermore, we also verify the validity of universal geometric relations for far-field modes, including the wave interference case,  Bessel and Laguerre-Gaussian beams with non-zero topological charges (The details can be found in Supplementary Information). 

\begin{figure*}[tp]
\centering
\includegraphics[width=\linewidth]{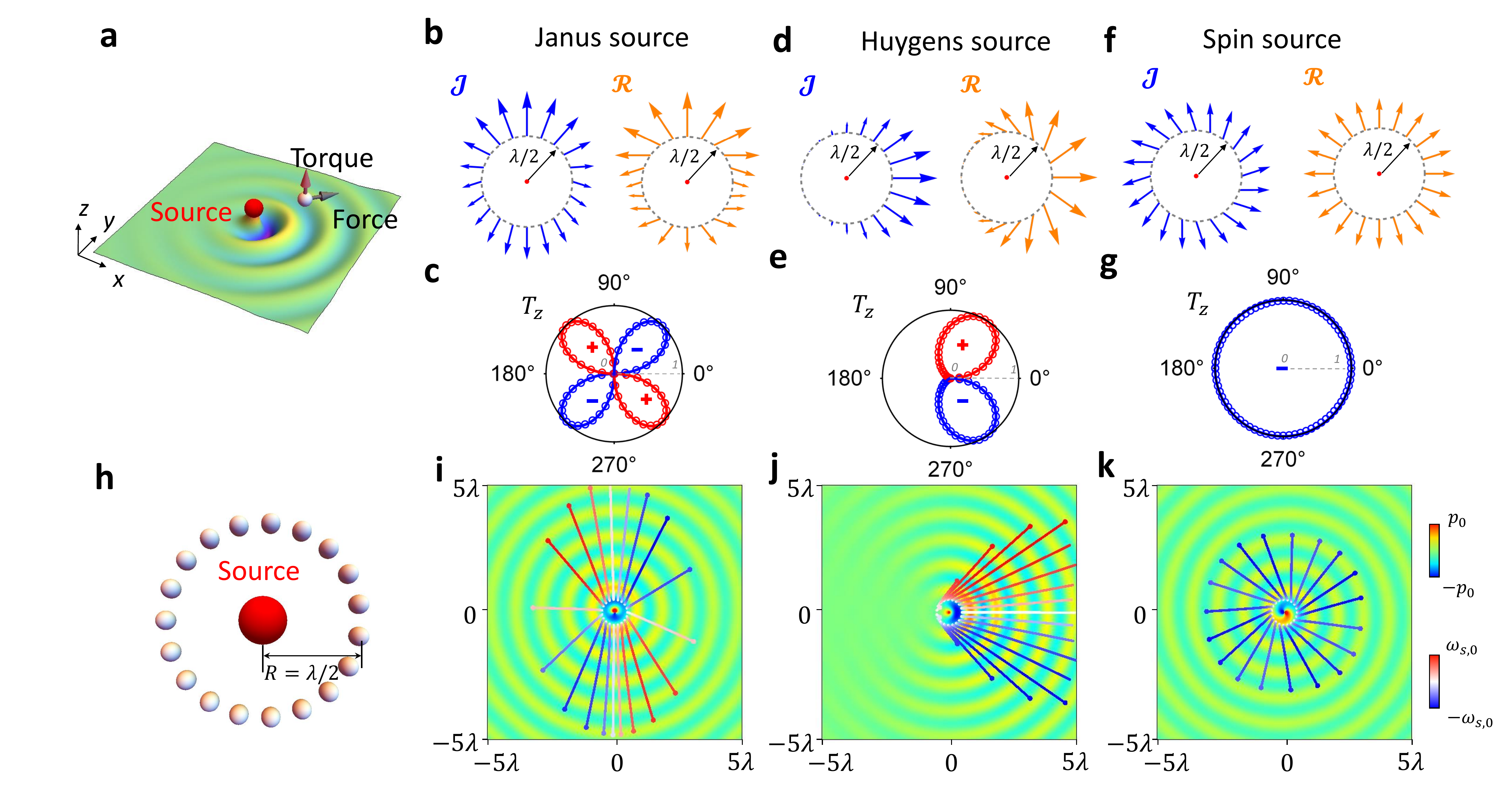}
\caption{ {\bf Mechanical interactions between the lossy small particle and near-field acoustic sources.} \textbf{a}, the acoustic source placed at the original point will apply non-zero force and torque on the small lossy particle with the radius $a$($ka = 0.1$). The distance between the source and particle is $\lambda/2$; \textbf{b,c} the $\bm{\mathcal{J}}$, $\bm{\mathcal{R}}$ and torque $T_z$ induced by the Janus source ($M+D_y$). The solid line and circles denote the analytical (Eq.~\ref{eq:dynamic}) and numerical acoustic spin torques, respectively. The radius means the absolute amplitude ($|T_z|$) and the color denotes the sign (red for positive, blue for negative). $T_z$ are normalized by the maximum of their absolute values; \textbf{d,e} the $\bm{\mathcal{J}}$, $\bm{\mathcal{R}}$  and torque $T_z$ induced by the Huygens source ($M+iD_x$); \textbf{f,g} the $\bm{\mathcal{J}}$, $\bm{\mathcal{R}}$  and torque $T_z$ induced by the Spin source ($D_x+iD_y$); \textbf{h}, the circle array of 19 particles is placed near the near-field sources with distance $R=\lambda/2$ initially. Their time evolutions (after $t = 1.5$s) are shown for: Janus source (\textbf{i}), Huygens source (\textbf{j}) and Spin source (\textbf{k}). The lines denote the particles' motion trajectories and their colors mean the angular frequency $\omega_s$ of the particles' self-rotations.  $m=\frac{4}{3} a^3 \rho_d$, $I_s = \frac{2}{5} m a^2$,  $\rho_d = 2 \rho$.}
\label{fig:torque}
\end{figure*}

{\bf Universal relations for acoustic torques and forces.}
We find that similar universal geometric relations also hold for the acoustic spin torque and acoustic forces for small absorptive particles~\cite{toftul2019acoustic, abdelaziz2020acoustokinetics, Marzo2018}. We can follow Ref.~\cite{toftul2019acoustic} to obtain the acoustic SAM absorbing induced local torque $\bm{T}$:
\begin{equation}
\bm{T} = \frac{4\rho}{|p|^2} {\rm Im}[\alpha_d] \bm{\mathcal{J}}\times\bm{\mathcal{R}}= \frac{{\rm Im}[\alpha_d]}{\omega}  \nabla\times\bm{\mathcal{J}},
\label{eq:dynamic}
\end{equation}
and the radiation force $\bm{F} = \bm{F}^{\rm grad} + \bm{F}^{\rm scat}$, as
\begin{eqnarray}
\bm{F}^{\rm grad} &=& -\frac{\omega}{c^2} ({\rm Re}[\alpha_m] + {\rm Re}[\alpha_d])\bm{\mathcal{R}} - \frac{1}{2\omega}{\rm Re}[\alpha_d]\nabla^2\bm{\mathcal{R}},
\\
\bm{F}^{\rm scat} &=& \frac{\omega}{c^2}({\rm Im}[\alpha_m] + {\rm Im}[\alpha_d])\bm{\mathcal{J}} + \frac{1}{2\omega}{\rm Im}[\alpha_d]\nabla^2\bm{\mathcal{J}}, 
\label{eq:dynamic2}
\end{eqnarray}
where $\alpha_m$ and $\alpha_d$ are the monopole and dipole polarizabilities of the small lossy particle~\cite{toftul2019acoustic}. $\bm{F}^{\rm grad}$ is the gradient part of total force with no curl ($\nabla\times\bm{F}^{\rm grad}=0$) and $\bm{F}^{\rm scat}$ is the scattering part with no divergence ($\nabla\cdot\bm{F}^{\rm scat}=0$).
As a consequence of Eq.~(\ref{eq:s}), for a small absorptive particle, we can obtain the following geometric orthogonal constraints for the SAM torque and acoustic forces (See Supplementary Information):
\begin{equation}
\begin{aligned}
\bm{T}\cdot\bm{F}^{\rm scat} &= \frac{1}{2} {\rm Im}[\alpha_d]^2 \bm{s}\cdot(\nabla^2\bm{\mathcal{J}}), \\
\bm{T}\cdot\bm{F}^{\rm grad} &= -\frac{1}{2}{\rm Re}[\alpha_d]{\rm Im}[\alpha_d] \bm{s}\cdot(\nabla^2\bm{\mathcal{R}}). 
\end{aligned}
\label{eq:orthogonal}
\end{equation}
It is clear that the geometric orthogonal condition between radiation forces and the SAM torque depends on whether $\bm{s}$ is orthogonal to $\nabla^2\bm{\mathcal{R}}$ and $\nabla^2\bm{\mathcal{J}}$, or whether $\nabla^2\bm{\mathcal{R}}$ and $\nabla^2\bm{\mathcal{J}}$ are zero. For the surface evanescent wave, one can find $\bm{T} \perp \bm{F}^{\rm scat}$, $\bm{T} \perp \bm{F}^{\rm gard}$, coinciding with previous predictions~\cite{toftul2019acoustic, Bruus2012}. Generally, from Eq.~(\ref{eq:orthogonal}),  the orthogonality between the SAM torque and the acoustic force is not guaranteed.

Now let us demonstrate the mechanical operations on small lossy particles induced by selective near-field sources. The excited pressure field of sources combined linearly by acoustic monopole $M$ and dipoles $\bm{D}$ can be written as: $p(\bm{r}) = i \rho \omega (M G(\bm{r}) - \frac{1}{k} \bm{D}\cdot \nabla G(\bm{r}))$, where $G(\bm{r}) = \frac{e^{i k r}} {4\pi r}$ is the Green function that the source is at the original point, $r = |\bm{r}|$ and $k=k_0$. For two-dimension case, the Green function will be $G(\bm{r}) = H_0^{(1)}(k r)$, where $H_0^{(1)}$ is the Hankel function of the first kind. 
As shown in Fig.~\ref{fig:torque}(a), the dynamics of a small lossy particle driven by acoustic fields can be described as: $m\frac{d^2\bm{r}}{dt^2}=\bm{F}$ and $I_s \frac{d \bm{\omega}_s}{dt}=\bm{T}$, where $m$ is the mass and $\bm{r}$ is the position of the particle, $I_s$ is the moment of inertia and $\bm{\omega}_s$ is the angular frequency along the particle's own axis. We consider that the particle radius $a$ is so small ($ka = 0.1$) that the back-action effect from the particle to wave fields is negligible. Specially, according to Eq.~(\ref{eq:orthogonal}), for near-field sources in 2D planes, their near-field waves will induce the geometrically orthogonal radiation forces and spin torques, $\bm{T} \perp \bm{F}^{\rm scat}$ and $\bm{T} \perp \bm{F}^{\rm gard}$, namely, the forces are in plane while the torques are out of plane.

Here, we consider three typical near-field sources placed at the origin point of the $xOy$ 2D dimension: acoustic Janus, Huygens and Spin sources~\cite{long2020symmetry}. The $\bm{\mathcal{J}}$, $\bm{\mathcal{R}}$ and the SAM torque along the $z$-axis $T_z$ in the near-field region at the distance $\lambda/2$ far away from the source are calculated in Fig.~\ref{fig:torque}. 
Without loss of generality, we consider the lossy particle with the relative density $ \overline{\rho}/\rho = 2 + 0.5i$ and compressibility $ \overline{\beta}/\beta = 3 + 0.7i$~\cite{toftul2019acoustic}.
For {\it Janus source} $M+D_y$ in Fig.~\ref{fig:torque}(b,c), the near-field directionality mainly comes from the asymmetric $\bm{\mathcal{R}}$. Combining with the slightly asymmetric $\bm{\mathcal{J}}$, Janus source produces the quadrupole-like spin torque: $T_z(\theta) = -T_z(\theta+\pi/2)$, where $\theta = {\rm arg}[x+iy]$ and ${\rm arg}[\cdot]$ means the angle of the complex. For {\it Huygens source} $M+i D_x$ in Fig.~\ref{fig:torque}(d,e), it has both the directional $\bm{\mathcal{J}}$ and $\bm{\mathcal{R}}$, producing the dipole-like SAM torque: $T_z(\theta)=-T_z(-\theta)$. For {\it Spin source} $D_x + i D_y$ in Fig.~\ref{fig:torque}(f,g), it induces the isotropic radiation pattern and monopole-like SAM torque $T_z(\theta) = {\rm const.} <0$. As verified, the universal geometric relations Eqs.~(\ref{eq:s},~\ref{eq:sJ},~\ref{eq:dynamic}) are generally valid in these three near-field source systems.

To demonstrate the effect of acoustic torques and radiation forces, we place $19$ particles around these sources with the distance $R=\lambda/2$ initially, and calculate their motions and trajectories, as illustrated in Fig.~\ref{fig:torque}(i,j,k): (i) Janus source; (j) Huygens source; (k) Spin source. From the simulations, one can see that the particles' trace and self-rotation angular frequencies $\omega_{s,z}$ reflect the mechanical operations of different sources: quadrupole, dipole and monopole-like SAM torque patterns. The details about radiation forces of these sources can be found in Supplementary Information. 

\begin{figure}[tp!]
\centering
\includegraphics[width=\linewidth]{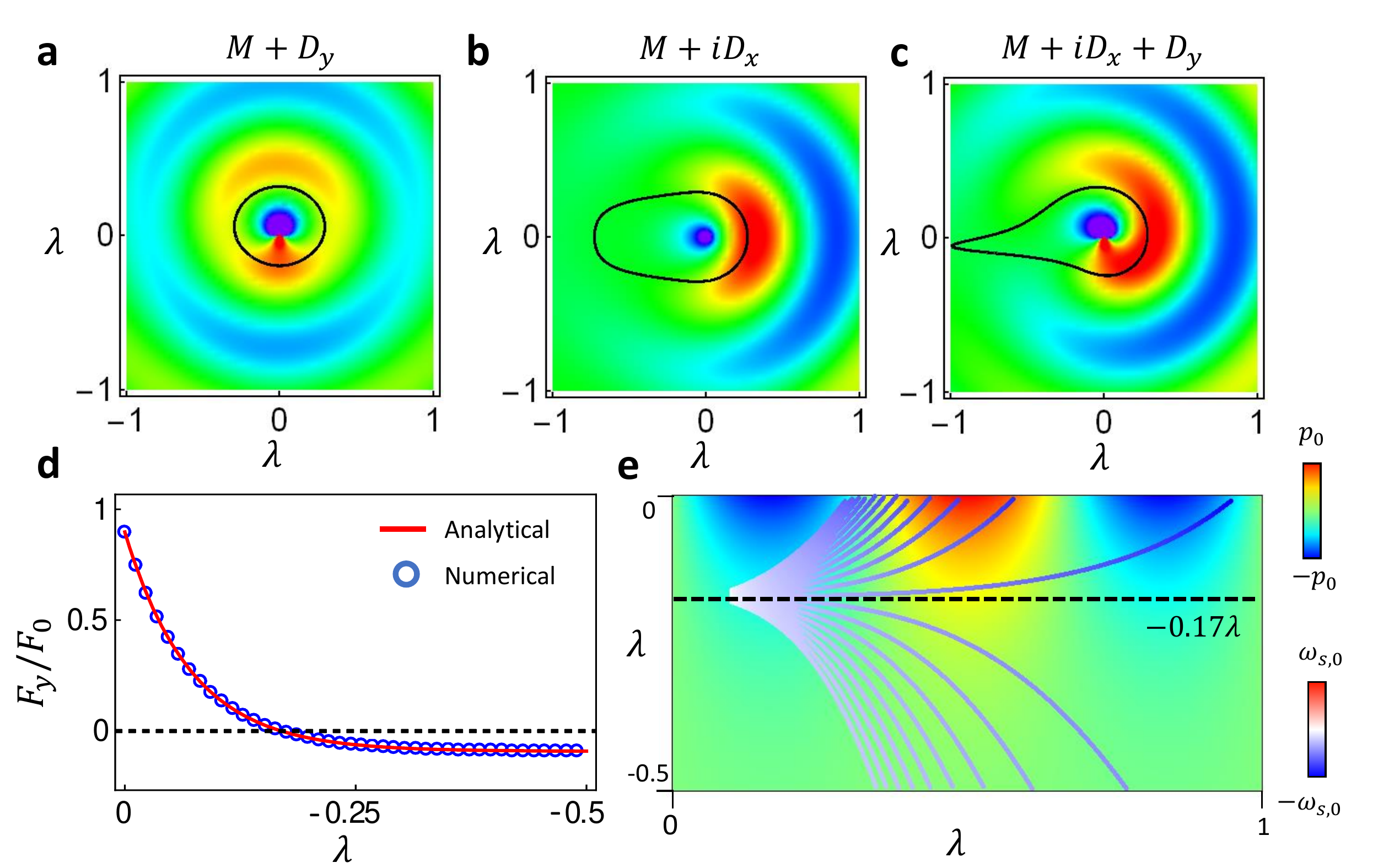}
\caption{{\bf The acoustic pulling for the small lossy particle ($ka\ll1$)}. \textbf{a,b,c} For the near-field sources, they will trap particle at the different positions (denoted as the black line) due to their characteristic near-field properties: Janus source $M+D_y$ (\textbf{a}), Huygens source $M+iD_x$ (\textbf{b}) and one combination $M+iD_x+D_y$ (\textbf{c}). \textbf{d} The relation between $F_y$ induced by the evanescent wave and the distance  of particle away from the interface. Obviously, the particle will be trapped at $\approx -0.17\lambda$. \textbf{e}, the time evolution (after $t=2$s) of the motion of particles placed around the distance $=0.17\lambda$ away from the interface. Here, $|p_0| = 400$ Pa, $k=1.5 k_0$, $F_0=|p_0|^2 a^2 \beta$, $F_b = 0$ for the simplicity. }
\label{fig:pulling}
\end{figure}

Besides pushing the particle, the near-field sources can also be exploited to pull the particle. As we will see, when the particle is closing to the center of near-field sources, the radiation pushing force may not be increasing, but becomes decreasing and even changes to an attracting force. As shown in Fig.~\ref{fig:pulling}(a), in the region denoted by elliptical black line, the acoustic Janus source will pull the particle to the source center, namely $F_n <0$. For Huygens source in Fig.~\ref{fig:pulling}(b), the pulling region will be deformed and become non-symmetric horizontally due to its relation with $\bm{\mathcal{J}}$~\cite{long2020symmetry}.  Specially, the pulling region of the near-field source can be manipulated by adjusting the linear combinations of acoustic monopole and dipoles, as shown in Fig.~\ref{fig:pulling}(c). Therefore, the black line plays a reminiscent role of the event horizon that particles within the critical region will collapse into the center. The conventional surface acoustic wave $p=p_0 e^{\tau y}e^{ikx}$, $y\in (-\infty, 0]$ can also be used to pull the particle along the $y$ axis even considering the gravity along the $-y$ and the buoyancy force $F_b$. Theoretically, there exists a critical condition for the evanescent wave to pull the particles (See Supplementary Information):
\begin{equation}
|p_0|^2 \geqslant \frac{2\rho \omega^2 (m g - F_b)}{2\tau^3 {\rm Re}[\alpha_d] + \tau k_0^2 {\rm Re}[\alpha_d + \alpha_m]}
\end{equation}
where $g$ is the gravitational acceleration. As an example as shown in Fig.~\ref{fig:pulling}(d), when the evanescent wave satisfies this condition, there is a region $y>-0.17\lambda$ where the total force compose of acoustic force and gravity points along $y$: $F_y >0$ so that the evanescent wave will pull the particles to the surface. Based on the numerical simulations, we can see that the particles released around $y=-0.17\lambda$ will have different traces: the particles within distance $-0.17\lambda$ will be pulled to the surface and the other particles will fall due to the gravity. Specially, due to spin-momentum locking, all particles will have negative self-rotation angular frequencies $\omega_{s,z} <0$.


To summarize, we have uncovered universal geometric relations of acoustic spin and radiation torque with energy flow and reactive power in acoustics. The intrinsic relations among. these acoustic quantities are. independent on wave structure details. The acoustic SAM is generally related the cross product of time-averaged energy flow and reactive power, as well as the local vorticity of energy flow. Similar geometric relations have also been uncovered for acoustic torques and forces, resulting in the geometrically orthogonal conditions. 
Spin-related mechanical behaviours of different near-field selective sources, Janus, Huygens and Spin sources have been demonstrated, exhibiting the quadrupole, dipole, monopole-like spin torque behaviors, respectively. Our work shows basic principles for physical mechanisms behind the acoustic SAM and would inspire new applications for SAM-related wave controls~\cite{Zhang2020, Long2020a, Yuan2021}, near-field directional excitations~\cite{long2020symmetry, Shi2021} and mechanical operations for small particles~\cite{Ding2013, Tian2020, Shen2018}. Our work would also be meaningful for other wave field systems including optical~\cite{Shi2021} and elastic waves~\cite{Yuan2021}.

~\\
\textbf{Data availability}
The data that support the findings of this study are available from the corresponding author upon reasonable request.

~\\
\textbf{Acknowledgments}
This work is supported by the National Natural Science Foundation of China (Nos. 11935010 and 11775159) and the Opening Project of Shanghai Key Laboratory of Special Artificial Microstructure Materials and Technology.

~\\
\textbf{Author Contributions}
Author contributions Y.L and J.R derived the the- ory and developed the analysis. Y.L and C.Y carried out the numerical simulations. J.R and H.C conceived the project. All the authors contributed to discussion, interpreting the data and the writing.

~\\
\textbf{Competing interests}
The authors declare no competing interests.

\bibliography{references}

\end{document}